**Surface geometry determined temperature-dependent band structure evolutions in organic halide perovskite single crystals**

Jinpeng Yang[1,2*], Matthias Meissner[2], Takuma Yamaguchi[2], Bin Xi[1], Keishi Takahashi[3], Shed Abdullah[3], Xianjie Liu[4], Hiroyuki Yoshida[3], Mats Fahlman[4], and Satoshi Kera[2*]

[1]College of Physical Science and Technology, Yangzhou University, Jiangsu, China

[2]Institute for Molecular Science, Department of Photo-Molecular Science, Myodaiji, Okazaki, Japan

[3]Graduate School of Engineering, Chiba University, Chiba, Japan

[4]Laboratory for Organic Electronics, ITN, Linköping University, Norrköping, Sweden

*Correspondence to: yangjp@yzu.edu.cn; kera@ims.ac.jp

**Abstract**

In this study, different electronic structure evolutions of perovskite single crystals are found via angle-resolved photoelectron spectroscopy (ARPES): (i) unchanged top valence band (VB) dispersions under different temperatures can be found in the $CH_3NH_3PbI_3$, (ii) phase transitions induced the evolution of top VB dispersions, and even a top VB splitting with Rashba effects can be observed in the $CH_3NH_3PbBr_3$. Combined with low-energy electron diffraction (LEED), metastable atom electron spectroscopy (MAES), and DFT calculation, we confirm different band structure evolutions observed in these two perovskite single crystals are originated from the cleaved top surface layers, where the different surface geometries with $CH_3NH_3^+$-I in $CH_3NH_3PbI_3$ and Pb-Br in $CH_3NH_3PbBr_3$ are responsible for finding band dispersion change and appearing of the Rashba-type splitting. Such findings suggest that the top surface layer in organic halide perovskites should be carefully considered to create functional interfaces for developing perovskite devices.

Organic-inorganic halide perovskites ($ABX_3$, where A = organic cation, B = Ge, Sn, or Pb, and X = halide ion) have several advantages that facilitate application in optoelectronics, such as long carrier lifetime, large diffusion length, high light absorption coefficient, low exciton binding energy, and easy fabrication. [1–5] As a result, perovskite solar cells are in further development toward commercialization despite a lack of fundamental understanding of several crucial areas. It has reported that electronic structure at surfaces of perovskite layers could strongly affect charge transport, device performance, and stabilities. [6-8] However, the physical understanding of electronic structure in surfaces of perovskites is still inconclusive. [9-11] Structure study has recently reported that the coexistence and gradual phase transition in these organic halide perovskites, while the mysterious inconsistency of observed band dispersion and surface structure is still waiting for urgent clarifications. [9, 11-13] On the other hand, Niesner et al. have found the Rashba-type effect induced top valence band (VB) splitting via angle-resolved photoelectron spectroscopy (ARPES) for $CH_3NH_3PbBr_3$ single crystal under both cubic and tetragonal phases, [14] which is later thought due to the surface polar domains and surface defects during sample preparation by Che et al. [15] In order to address these inconsistencies and controversies, a direct study of temperature-induced phase transitions at surface structures as well as evolutions of their electronic structures is then essential, which unfortunately still has remained missing.

In this report, we use ARPES, low-energy electron diffraction (LEED), and metastable atom electron spectroscopy (MAES) techniques to study the evolution of electronic structures and the change in surface geometry at various temperatures of the organic halide perovskite single crystals. We identify different characters of band dispersions between $CH_3NH_3PbBr_3$ and $CH_3NH_3PbI_3$ (A = $CH_3NH_3^+$, B = Pb, and X = Br or I), even though their LEED results yield similar patterns. In contrast to no change of band structures in $CH_3NH_3PbI_3$ single crystal during the decreasing of temperatures, clear band structure evolutions with gradually appearing the top VB splitting can be found in the $CH_3NH_3PbBr_3$ single crystal. Based on additional experimental evidence from LEED and MAES, it is found that different surface geometries cleaved from organic halide perovskites (A-X plane and B-X plane) dominate these various observations in band structures as well as the origin of top VB splitting due to the Rashba-type effect, which is strongly suggesting us the importance of preparing different top surface layer in organic halide perovskites for developing new functional devices.

Organic halide perovskite single crystals were grown by the following methods proposed by Poglitsch and Weber, [16] in which their crystal structures were confirmed using crystal X-ray diffraction (supplementary information **Table S1**). All cleaved samples were adhered to molybdenum sample holders by using a silver paste in the atmosphere for a short time (~2 min), followed by annealing in an ultrahigh-vacuum ($<3\times10^{-8}$ Pa), while the temperature slowly increased to 350 K (1 h) to confirm surface cleanness. The ARPES was performed using a high-sensitivity apparatus with a hemispherical electron energy analyzer (MBS A-1), monochromatic He $I_\alpha$ (hν = 21.22 eV) radiation source, where the resolution was set to 30meV. Temperature-dependent

LEED measurements were performed in a calibrated microchannel-plate (MCP) LEED device attached to the ARPES UHV chamber. MAES measurements were performed in Chiba University, where metastable He* atoms ($2^3$S: 19.82 eV) were generated by a cold cathode discharging, and He*($2^1$S) was eliminated by using a quench lamp (DC helium lamp). The DFT calculations were performed with Perdew–Burke–Ernzerhof (PBE) generalized gradient approximation (GGA) including spin-orbit coupling (SOC) in VASP.

**Fig. 1** depicts the ARPES spectra of (a) $CH_3NH_3PbI_3$ and (b) $CH_3NH_3PbBr_3$ with decreasing temperatures along the cubic ΓM ($Γ^CM^C$) direction. The second-derivative spectra are also provided to enhance visibility, which can be seen in **Fig. S1**. As shown in **Fig. 1(a)**, clear band dispersions of the top VB along the $Γ^CM^C$ direction for $CH_3NH_3PbI_3$ are obtained under 350 K, in which the top VB near the $Γ^C$ point can be seen with a flat band. In contrast, an apparent change in the top VB at the $M^C$ point is found with a dispersion period of 2×0.70 Å$^{-1}$ and a total energy shift of ~0.7 eV. These observations agree with calculations that only incorporate a cubic phase [9] Moreover, when measured temperatures become much lower (from 300 K to 100 K), the observed dispersion does not demonstrate any change both in the periodicity and in the energy position, which seems to indicate the unchanged surface geometry and related electronic structures. The effective mass can be estimated from these spectra by using the following equation: $m^*(k) = \hbar^2 \times (\partial^2 E(k)/\partial k^2)^{-1}$, where $m^*(k)$, $\hbar$, $E(k)$ and $k$ are the effective mass, reduced Planck's constant, energy, and wave vector, respectively.

The m*= (0.24±0.05) $m_0$ ($m_0$ is the free-electron mass) obtained without any apparent change from temperature reduction.

On the other hand, the ARPES spectra for $CH_3NH_3PbBr_3$ (in **Fig. 1(b)**) demonstrate remarkable changes in the band dispersions on the top VB compared with results from $CH_3NH_3PbI_3$. At 300 K, we observe a clear band dispersion of the top VB with a maximum at the $M^C$ point (~0.75 Å$^{-1}$), consistent with previously reported results. [10] Furthermore, a gradual evolution of the band dispersion on the top VB is seen when the temperature decreases from 300 K to 100 K. An additional top VB appears at the Γ point and can be distinguished at both 150 K and 100 K, thus indicating that the dispersion period changes from ~2×0.75 Å$^{-1}$ to ~2×0.38 Å$^{-1}$. More importantly, in the spectrum of a $CH_3NH_3PbBr_3$ sample measured at 100 K, the top VB at the two Γ points ($k_{//}$ = 0 Å$^{-1}$ and ~0.75 Å$^{-1}$) demonstrates a significant splitting-like behavior, confirms the Rashba-type effects proposed by many theorists over the years. [15, 17-21] The m*= (0.26±0.05) $m_0$ obtained is consistent with a previous report (see detailed fitting in **Fig. S2**). [10] Meanwhile, the Rashba parameter, $α_R=2E_R/k_0$ is estimated to be 3±1 eV·Å (with $k_0$ = 0.1 Å$^{-1}$ and $E_R$ = 0.15 eV). The VB dispersions for $CH_3NH_3PbI_3$ and $CH_3NH_3PbBr_3$ along the $Γ^CX^C$ (a cubic phase) direction measured at two different temperatures are also shown in **Fig. S3.** Different band structure evolutions can be also seen between $CH_3NH_3PbI_3$ and $CH_3NH_3PbBr_3$.

The different top VB change between $CH_3NH_3PbI_3$ and $CH_3NH_3PbBr_3$ with reducing the temperatures seems to imply the different surface structures, which is initially studied with LEED measurements. As shown in **Fig. 2**, the distortion-corrected

LEED results are provided with measured temperatures changing from 350 K to 100 K for $CH_3NH_3PbI_3$ and from 300 K to 100 K for $CH_3NH_3PbBr_3$. The simulated structures with necessary parameters are also superimposed on the half side of each image. In **Fig. 2 (a)**, we observe a gradual change in the surface structure with lattice constants of square unit cells varying from 6.25±0.02 Å (white color, both at 350 K and 300 K), 8.86±0.02 Å (green color, at 250 K) to 12.53±0.02 Å (yellow color, both at 150 K and 100 K) for the $CH_3NH_3PbI_3$ single crystal, which agrees with the previously reported results with phase changes from cubic, tetragonal to orthorhombic. [16] The gradual phase transition induces mixed phases that can also be observed in our LEED results marked with red arrows. Weak LEED spots representing a tetragonal phase is found at 300 K and an orthorhombic phase is evident at 150 K. Furthermore, the inconsistency is obviously shown between results from LEED and ARPES since no change of band dispersion in the periodicity is found from ARPES (**Fig. 1 (a)**). In **Fig. 2(b)**, the surface structure for the $CH_3NH_3PbBr_3$ gives only two different lattice constants of square unit cells among different temperatures, in which one is of 5.93±0.05Å (white color, 300 K, representing a cubic phase) and the other one is of 8.38±0.05Å (green color, from 250 K to 100 K, representing a tetragonal phase). Weaker spots representing a tetragonal phase at 300 K can be also found (marked with a red arrow). Different from the inconsistent tendency in $CH_3NH_3PbI_3$, a good agreement can be found in $CH_3NH_3PbBr_3$. The periodic change in LEED results is consistent with ARPES results, where the surface structure changes from a cubic phase to a tetragonal phase.

Moreover, the surface structure has also been further studied by using MAES. Since MAES uses metastable atoms which cannot penetrate the bulk of the solid, it can enable the observation of the outermost surface electrons selectively. [22] Experimental results of MAES for crystal surfaces are shown in **Fig. 3**. The spectral features are assigned according to the DFT calculation, where the calculated spectra depending on the element are shown on the bottom of **Fig. 3**. In **Fig. 3(a)**, the MAES features keep almost unchanged for an intense top band which is described from I(5p) orbital upon the cooling, while increases for weak-Pb(6p)-related bands which are enhanced at the lower temperatures by displacing the surface I atoms. In **Fig. 3(b)**, the MAES features show no clear difference on the temperature, suggesting the orbital distribution at the outermost surface keeps unchanged even if the lateral displacement has occurred upon the cooling. These results demonstrate that the crystal surface should be composed by different geometries/elements, where the plane A-X ($CH_3NH_3^+$-I) should be for $CH_3NH_3PbI_3$ while plane B-X (Pb-Br) should be for $CH_3NH_3PbBr_3$.

Finally, based on our experimental observations combined with the theoretical calculations, the origin of the different evolution in ARPES spectra between $CH_3NH_3PbI_3$ and $CH_3NH_3PbBr_3$ can be elucidated. **Fig. 4** shows real space representations of surface geometries and the top VB wave functions of (a) a cubic $CH_3NH_3PbI_3$ and (b) a tetragonal $CH_3NH_3PbBr_3$, which depicts the assembly of the wave functions at atoms in different planes. Calculated band structure of $CH_3NH_3PbI_3$ and $CH_3NH_3PbBr_3$ with spin-orbital coupling are also given in **Fig. S4**. As shown in **Fig.**

**4**, the top VB wave functions are only assembled at the X (X=I or Br) atoms in the $CH_3NH_3^+$-X plane, either form a cubic phase or a tetragonal phase, suggesting the negligible influence of organic cation contributes to the density of states (DOS) of top VB. Moreover, combined with (i) the direct information on the atomic orbital spread at the outmost surface of single crystals given by MAES, and (ii) the phase transition from cubic to tetragonal mainly changes the arrangement of organic cations in $CH_3NH_3^+$-X plane (in the left of **Fig. 4**), it is then reasonable to expect that the unchanged band dispersions in the $CH_3NH_3PbI_3$ single crystal (**Fig. 1(a)**) are ascribed to the surface geometry with the $CH_3NH_3^+$-I plane. Despite the fact that temperature changes induced different phases (as confirmed from our LEED patterns in **Fig. 2**) in the $CH_3NH_3PbI_3$ single crystal, the change in the geometry on the $CH_3NH_3^+$-I plane is mainly focused on directions of the organic cations, such that the organic cations do not contribute any wave functions (or DOS) to top VB, which results in the inconsistency between the ARPES results and LEED results. On the other hand, the consistent experimental results of ARPES and LEED within a different surface geometry of the Pb-Br plane for the $CH_3NH_3PbBr_3$ single crystal should be achieved due to the broad distributions of the top VB wave functions (or DOS) in all surface atoms. The Pb-Br surface plane is also a key to realize the "experimentally visible" Rashba splitting via ARPES. Consequently, the origin is revealed from the ARPES results to confirm the existence of Rashba-type splitting in organic halide perovskites, not due to the commonly considered surface polar domains and surface defects. [15] The surface geometry/element dominates such unique observations not only in band structure evolution but also in spin degeneracies.

To summarize, we have successfully obtained the temperature-induced evolutions of the electronic structure in organic halide perovskite single crystals, where different characters of band dispersions between $CH_3NH_3PbBr_3$ and $CH_3NH_3PbI_3$ can be found, even though their LEED results yield similar patterns. Combined with MAES results and the DFT computations, these different observations can be well explained after considering the properties of the electronic structures at different crystal surfaces terminations (A-X plane and B-X plane). The Pb-Br plane on the surface of $CH_3NH_3PbBr_3$ dominates the observations of phase transitions from ARPES as well as the existence of Rashba-type splitting of top VB. In a tetragonal $CH_3NH_3PbBr_3$, we find that the Rashba-type splitting of the top VB in the momentum space is given by $k_0 = 0.1$ Å$^{-1}$ and $E_R = 0.15$ eV. More broadly, our findings here not only provide direct experimental proof of the novel electronic structure evolutions but also pave the way for extending organic halide perovskites from solar cells into spintronic devices.

This work is financially supported in part by JSPS KAKENHI (No. JP26248062, 18H03904), sponsored both by Qing-Lan Project from Yangzhou University, and China Scholarship Council. J. Yang and S. Kera designed the experiments. J. Yang, M. Meissner, T. Yamaguchi and X. Liu performed the experiments of LEED and ARPES. J. Yang, K. Takahashi, S. Abdullah performed the experiments of MAES. B. Xi were responsible for the DFT calculations. H. Yoshida and M. Fahlman contributed to content discussions. All the authors discussed the results and wrote the manuscript. Authors declare no competing interests. All data is available in the main text or the supplementary information.

**Fig. 1.** Band evolution of the organic halide perovskite single crystals with decreasing temperatures. (a) He I$_\alpha$-ARPES spectra of the CH$_3$NH$_3$PbI$_3$ single crystal along a cubic ΓM (Γ$^C$M$^C$) direction with temperature changes from 350 K to 100 K; (b) He I$\alpha$-ARPES spectra of the CH$_3$NH$_3$PbBr$_3$ single crystal along the Γ$^C$M$^C$ direction with temperatures changes from 300 K to 100 K.

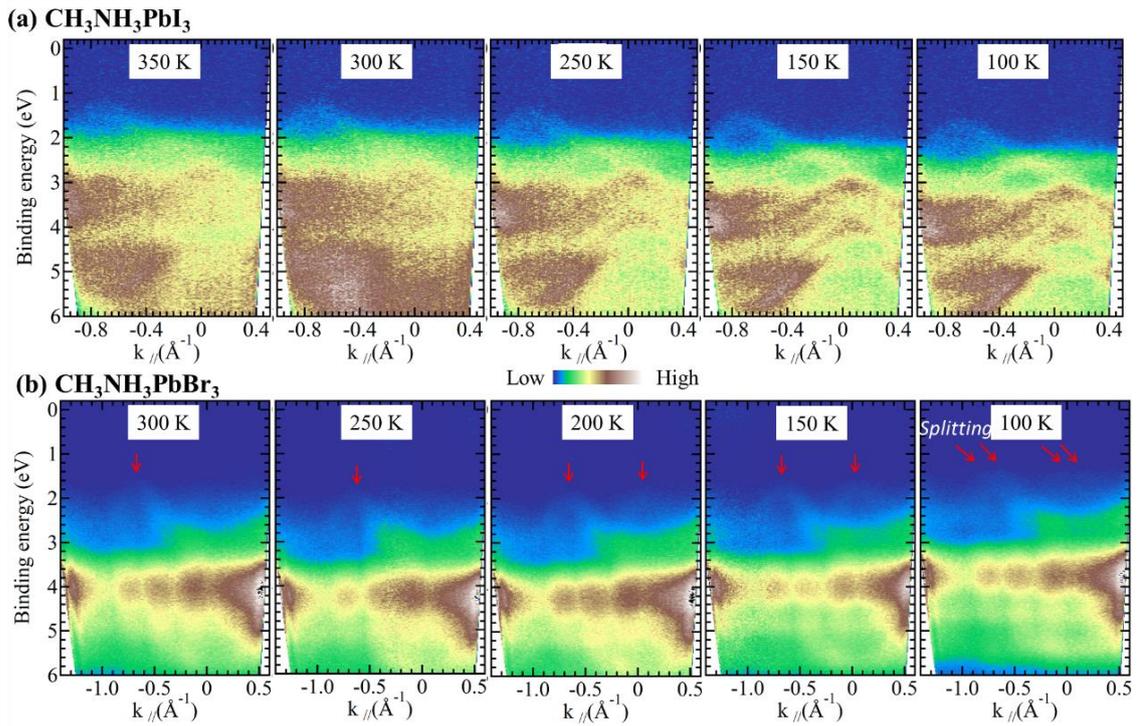

**Fig. 2.** LEED results of the distortion-corrected images of cleaved organic halide perovskites. (a) Surface of the CH$_3$NH$_3$PbI$_3$ single crystal, (b) surface of the CH$_3$NH$_3$PbBr$_3$ single crystal measured at various temperatures (as indicated). In CH$_3$NH$_3$PbI$_3$, the simulated structures having square unit cells with lattice constants of 6.25±0.02 Å (white), 8.86±0.02 Å (green), and 12.53±0.02 Å (yellow) are superimposed on the right side of each image. In CH$_3$NH$_3$PbBr$_3$, the simulated structures with lattice constants of 5.93±0.05 Å (white) and 8.38±0.05 Å (green) are superimposed on the left side of each image.

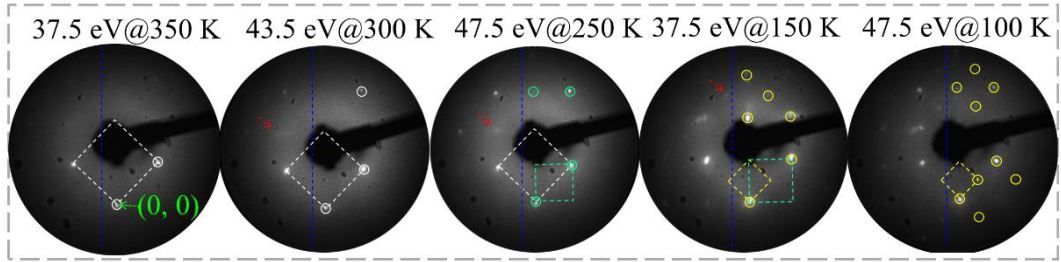

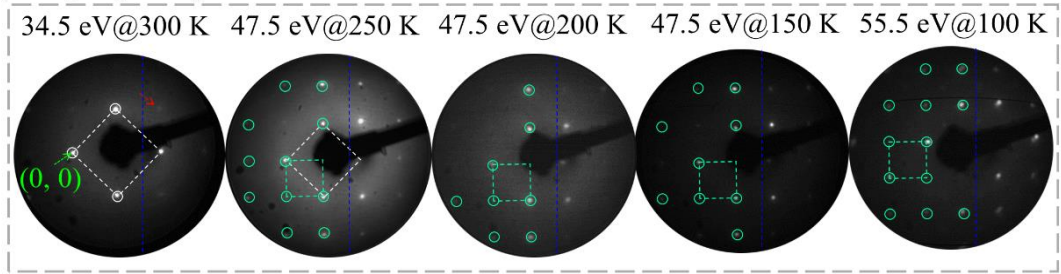
14

**Fig. 3.** Temperature-dependent MAES spectra obtained with temperature change from 300 K to 120 K for (a) CH$_3$NH$_3$PbI$_3$ and (b) CH$_3$NH$_3$PbBr$_3$ single crystals. Bottom spectra with gray color depict DFT calculated results with a tetragonal phase, where spectral features are assigned with different elements by using red arrows. The inset shows different top surface atomic structures for a tetragonal phase.

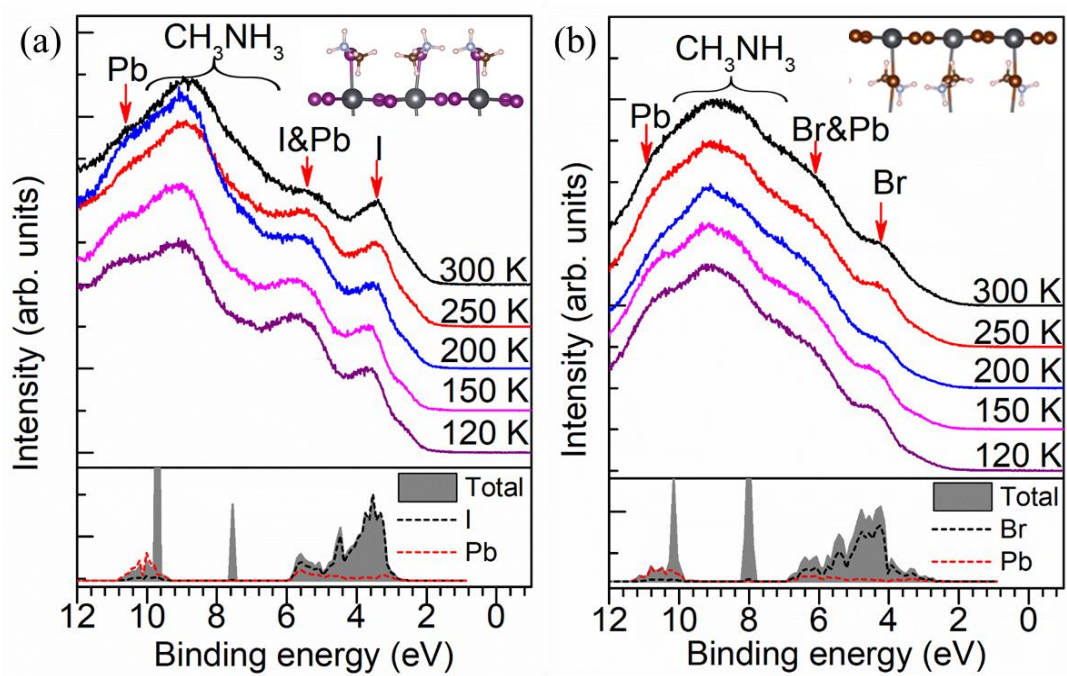

**Fig. 4.** A real space representation of the top VB wave functions of (a) a cubic CH$_3$NH$_3$PbI$_3$ single crystal near M point, and (b) a tetragonal CH$_3$NH$_3$PbBr$_3$ single crystal near Γ point, which depicts the assembly of the wave functions at atoms in two different planes (A-X plane and B-X plane, respectively). The left side of each figure shows the atomic arrangements in each plane.

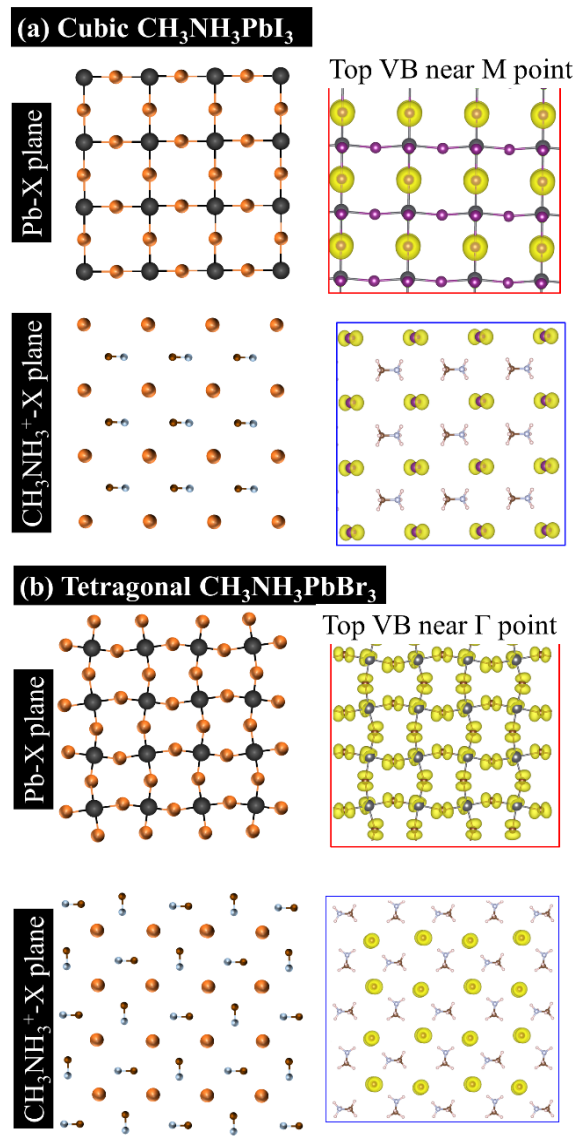

**Supplementary information for**

**Surface geometry determined temperature-dependent band structure evolutions in organic halide perovskite single crystals**

Jinpeng Yang[1, 2*], Matthias Meissner[2], Bin Xi[1], Keishi Takahashi[3], Shed Abdullah[3], Xianjie Liu[4], Hiroyuki Yoshida[3], Mats Fahlman[4], and Satoshi Kera[2*]

[1]College of Physical Science and Technology, Yangzhou University, Jiangsu, China

[2]Institute for Molecular Science, Department of Photo-Molecular Science, Myodaiji, Okazaki, Japan

[3]Graduate School of Engineering, Chiba University, Chiba, Japan

[4]Laboratory for Organic Electronics, ITN, Linköping University, Norrköping, Sweden

**Table S1.** Single crystal data for CH$_3$NH$_3$PbI$_3$ and CH$_3$NH$_3$PbBr$_3$.

| **Empirical Formula** | **CH6I3NPb (350K)** | **CH6Br3NPb (298K)** |
|---|---|---|
| **Formula Weight** | 619.98 | 478.98 |
| **Crystal Dimensions** | 0.200 × 0.200 × 0.200 mm | 0.200 × 0.200 × 0.200 mm |
| **Crystal System** | cubic | cubic |
| **Lattice Type** | Primitive | Primitive |
| **Lattice Parameters** | a = 6.29 Å | a = 5.90 Å |
|  | V = 248.86 Å$^3$ | V = 206.24 Å$^3$ |
| **Space Group** | P-43m (#215) | P-43m (#215) |
| **Z value** | 1 | 1 |
| **F000** | 260 | 260 |
| **m(MoKa)** | 261.937 cm$^{-1}$ | 261.937 cm$^{-1}$ |

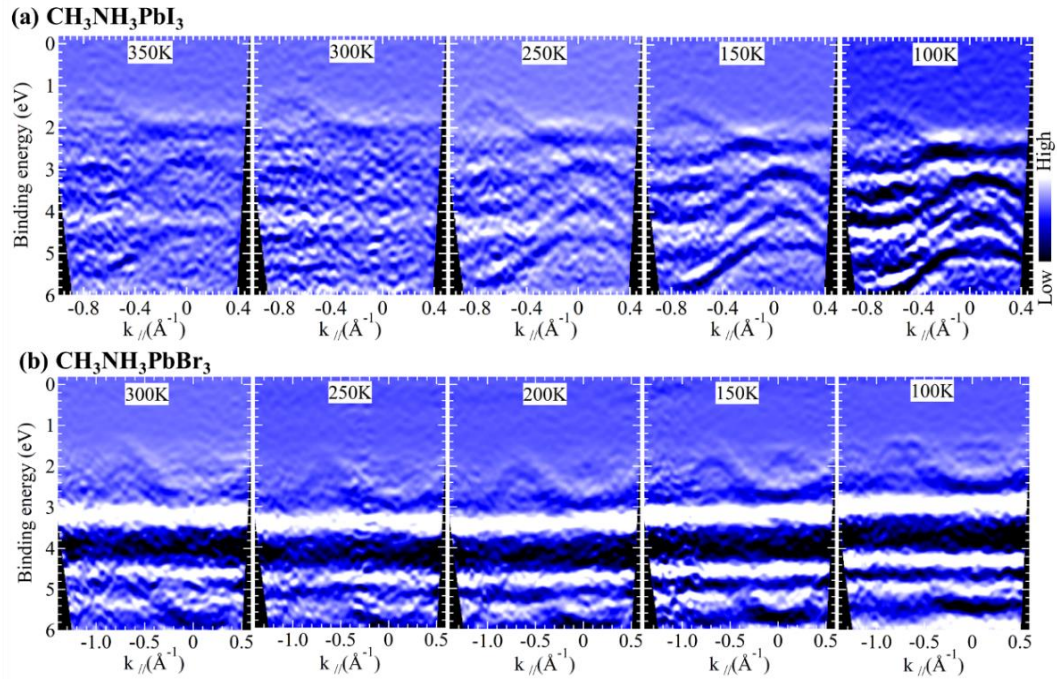

**Fig. S1.** ARPES second-derivative spectra of **Fig. 1** on band evolution with decreasing temperatures. (a) Spectra of CH$_3$NH$_3$PbI$_3$ single crystal along a cubic ΓM (Γ$^C$M$^C$) direction with temperatures change from 350 K to 100 K; (b) Spectra of CH$_3$NH$_3$PbBr$_3$ single crystal along Γ$^C$M$^C$ direction with temperatures change from 300 K to 100 K.

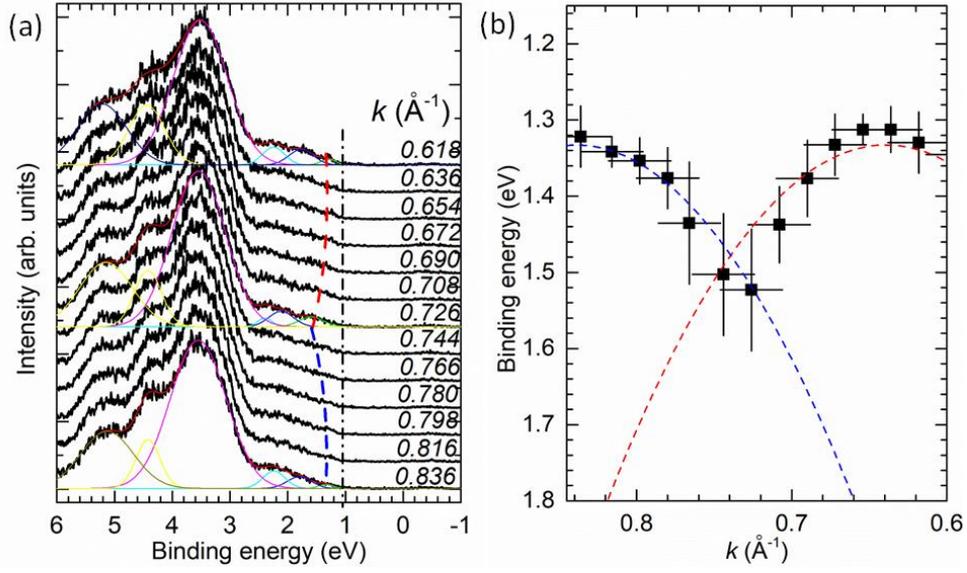

**Fig. S2.** (a) ARPES spectra for photoelectron emission momentum varying between 0.600~0.845 Å$^{-1}$ for demonstrating the Rashba splitting. The band dispersions are marked with the blue and red line as a guide to the eye. (b) Data fitting by using parabolic functions to estimate the effective mass and Rashba parameter.

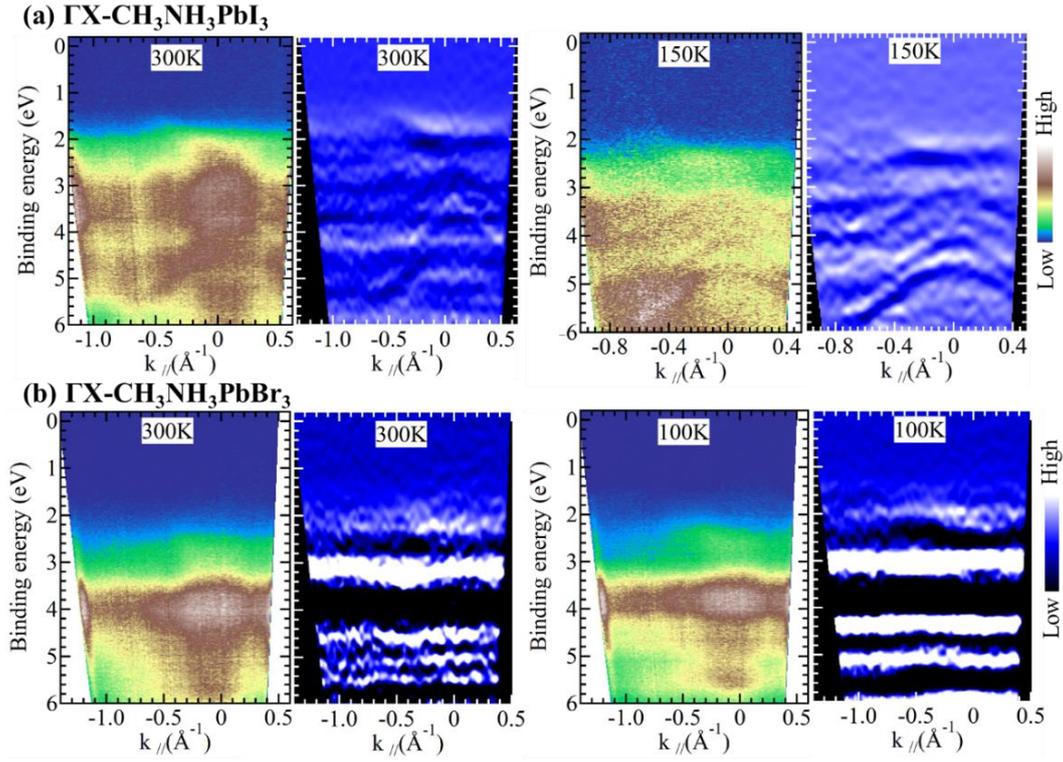

**Fig. S3.** ARPES along a cubic ΓX (Γ$^C$X$^C$) direction of (a) the CH$_3$NH$_3$PbI$_3$ single crystal and (b) the CH$_3$NH$_3$PbBr$_3$ single crystal under different temperatures. The raw spectra (left side) and -$d^2I(E)/dE^2$ spectra (right side) are both given for the $E$-$k_{//}$ dispersion.

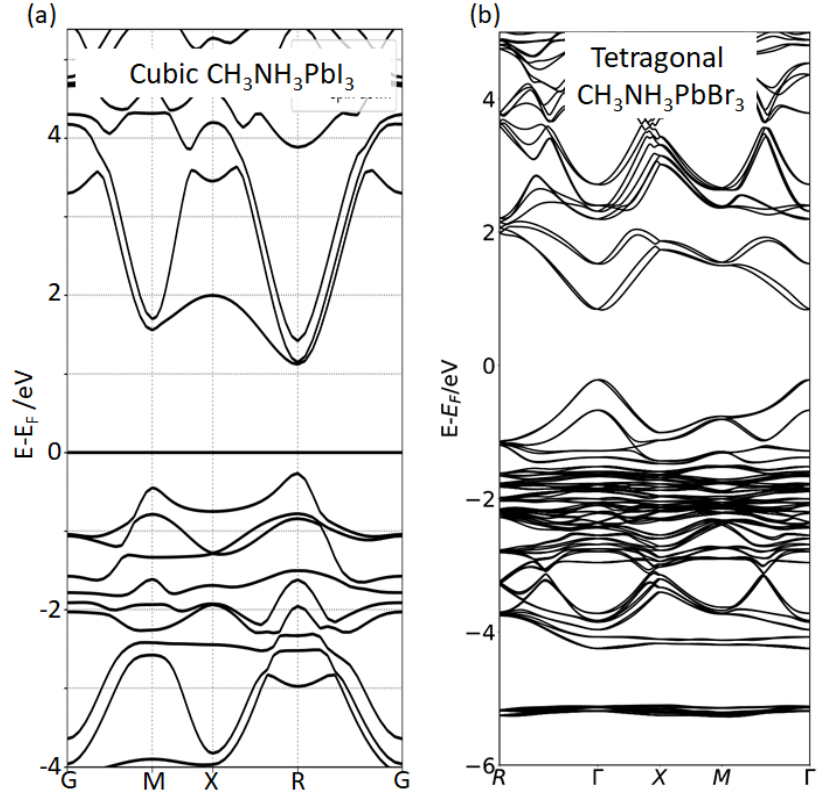

**Fig. S4.** Calculated band structures of (a) cubic CH₃NH₃PbI₃ and (b) tetragonal CH₃NH₃PbBr₃ with spin-orbital coupling. It is noticeable that the clear Rashba-type orbital splitting for both the top VB and bottom conduction band can only be found at the $\Gamma^T X^T$ region of tetragonal CH₃NH₃PbBr₃, while gave no orbital splitting in cubic CH₃NH₃PbI₃.